\documentclass[fleqn,twoside]{article}
\usepackage{espcrc2}
\usepackage{graphicx,epsfig,amssymb} 
\def\lsim{\raise0.3ex\hbox{$\;<$\kern-0.75em\raise-1.1ex\hbox{$\sim\;$}}}
\def\gsim{\raise0.3ex\hbox{$\;>$\kern-0.75em\raise-1.1ex\hbox{$\sim\;$}}}

\def\theta{\vartheta}

\newcommand{\beq}{\begin{equation}}
\newcommand{\eeq}{\end{equation}}
\newcommand{\ba}{\begin{eqnarray}}
\newcommand{\ea}{\end{eqnarray}}

\newcommand{\AmS}{{\protect\the\textfont2
  A\kern-.1667em\lower.5ex\hbox{M}\kern-.125emS}}

\title{Ultra High Energy Cosmic Ray Protons: Signatures and Observations
}

\author{V. Berezinsky\address
{ INFN, Laboratori Nazionali del Gran Sasso, I-67010 Assergi (AQ), Italy} 
}       
\begin{document}

\begin{abstract}
The  status of the Greisen-Zatsepin-Kuzmin (GZK) cutoff and pair-production dip
in Ultra High Energy Cosmic Rays (UHECR) is discussed. 
They are the features in the spectrum of protons propagating
through CMB radiation in extragalactic space, and discovery of these
features implies that primary particles are mostly extragalactic
protons. The  spectra measured 
by AGASA, Yakutsk, HiRes  and Auger detectors are in good agreement
with the pair-production dip, and HiRes data have strong evidences for  
the GZK cutoff. The Auger spectrum, as presented at the 30th ICRC 2007,
agrees with the GZK cutoff, too. The AGASA data agree well with the beginning 
of the GZK cutoff at $E \lsim 8\times 10^{19}$~eV, but show the excess 
of events at higher energies, the origin of which is not understood. 
The difference in the absolute fluxes measured by different detectors 
disappears after energy shift within the systematic errors of each
experiment. 

\vspace{1pc}
\end{abstract}

\maketitle

\section{Introduction}
\label{sec:introduction}
The systematic study of Ultra High Energy Cosmic Rays (UHECR)
started in late fifties after construction of Volcano Ranch (USA)
and Moscow University (USSR) arrays. 
At present due to the data of the last
generation arrays, AGASA, HiRes, and Pierre Auger observatory, 
 we are probably very close to understanding the origin of UHECR. 

The spectra of four detectors Yakutsk \cite{Yakutsk}, AGASA \cite{AGASA},
HiRes {\cite{Hires} and Auger \cite{Auger} are displayed in 
Fig.~\ref{fig:data}. One can see the great difference in the  
fluxes, but this difference is affected by a way of
presentation: The spectra are multiplied to $E^3$ and thus systematic
errors in energy determination strongly affect the displayed values. 
\begin{figure}[t]
\begin{center}
\mbox{\includegraphics[width=0.4\textwidth]{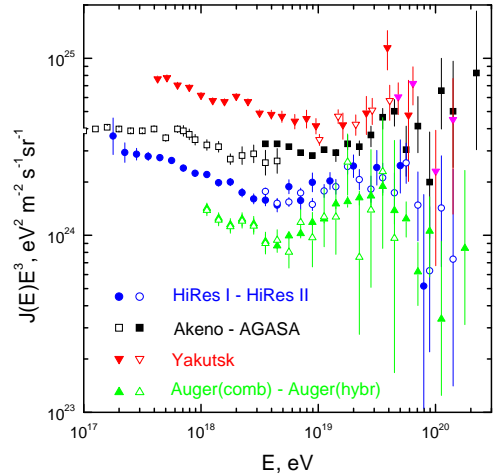}}
\end{center}
\caption{The spectra of UHECR measured by Yakutsk, AGASA, HiRes and 
Auger detectors.}
\label{fig:data}
\end{figure}

The nature of signal carriers of UHECR is not yet
established. The most natural primary particles are extragalactic
protons. Due to interaction with
the CMB radiation the UHE protons
from extragalactic sources are predicted to have a sharp
steepening of energy spectrum, so-called GZK cutoff \cite{GZK}.
It appears due to pion production in collisions of UHE protons
with CMB photons. 
Another signature of extragalactic protons in the primary spectrum is 
the dip \cite{BG88} - \cite{BGG-dip}. It is produced  due to
$p+\gamma_{\rm CMB} \to p+e^++e^-$ interaction with CMB. 
Being relatively faint feature, the dip is however clearly seen in the
spectra observed by Yakutsk, AGASA, Fly's Eye, 
HiRes, and Auger arrays.
This good agreement must be considered as a proof of a large fraction
of protons in the spectrum. 
The GZK cutoff, confirmed  by HiRes observations,
is also an evidence of the proton composition. 

The direct measurements of UHECR mass composition is contradicting. 
While HiRes data favour \cite{Hires-comp} the pure proton composition, 
Auger measurements indicate the mixed-nuclear composition \cite{Auger}.  

\section{Pair-production dip.}
\label{sec:thedip}
The analysis of the dip and GZK cutoff is convenient to perform in terms 
of the {\em modification factor}. It is defined as a ratio of the 
spectrum $J_p(E)$, calculated 
with all energy losses taken into account, and unmodified
spectrum $J_p^{\rm unm}$, where only adiabatic energy losses (red shift) are
included.
\beq  
\eta(E)=J_p(E)/J_p^{\rm unm}(E). 
\label{eq:modif}  
\eeq  

Modification factor is less model-dependent quantity than the
spectrum. In particular, it depends weakly on generation index 
$\gamma_g$, because
both numerator and denominator in Eq.~(\ref{eq:modif}) include
$E^{-\gamma_g}$. The dip and beginning of the GZK cutoff in  terms 
of the modification factor do not depend of distances between sources, 
different modes of proton propagation (from rectilinear to diffusion),
local overdensity and deficit of the sources etc (see \cite{BGG-prd}). 
 In Fig~\ref{fig:mfactor} the modification factors are shown 
for two spectrum indices $\gamma_g=2.0$ and $\gamma_g=2.7$.
They do not differ much. 

The dip in Fig.~\ref{fig:mfactor} has two 
flattenings. The low-energy flattening at $E \sim 1\times 10^{18}$~eV
provides transition to galactic cosmic rays, since the steep galactic
component ($\propto E^{-3.1}$) unavoidably intersects the flat 
extragalactic spectrum at $E \lsim 1\times 10^{18}$~eV. The high-energy
flattening explains {\em ankle}
observed at  $E \sim 1\times 10^{19}$~eV. 

Comparison of the predicted dip ($\eta_{ee}$ curve)  with observational 
data are shown in Fig~\ref{fig:dips}. For comparison $\gamma_g$ has to
be fixed. The values between 2.6 and 2.7 provides good agreement
of experimental and  theoretical curves. However, one can see  the
disagreement of the Akeno-AGASA and HiRes modification factors with 
theoretical prediction at $E \lsim 1\times 10^{18}$~eV 
(see Fig.~\ref{fig:dips}). By definition 
the modification factor (\ref{eq:modif}) cannot exceed unity, while  
the Akeno-AGASA and HiRes modification factors exceed this bound at 
$E \lsim 1\times 10^{18}$~eV. It 
signals the appearance of the new component of cosmic rays at 
$E \lsim 1\times 10^{18}$~eV and this component can be nothing, but
galactic cosmic rays. For the detailed discussion of the transition   
see \cite{BGG-prd} and \cite{Aloisioetal}.
\begin{figure}[ht]
\begin{center}
\mbox{\includegraphics[width=0.4\textwidth]{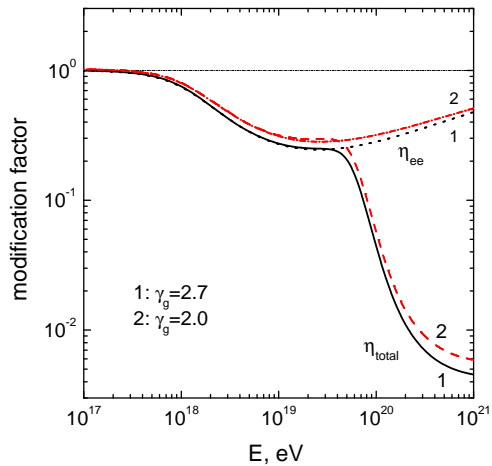}}
\end{center}
\caption{Modification factor for the power-law generation spectra
with $\gamma_g$ in the range 2.0 - 2.7. Curve $\eta=1$ corresponds to
adiabatic energy losses, curves $\eta_{ee}$ - to adiabatic and pair
production energy losses and curves $\eta_{tot}$ - to total energy losses.}
\label{fig:mfactor}
\end{figure}

\begin{figure*}[t]
  \begin{center}
  \mbox{\includegraphics[height=5cm,width=7.5cm]{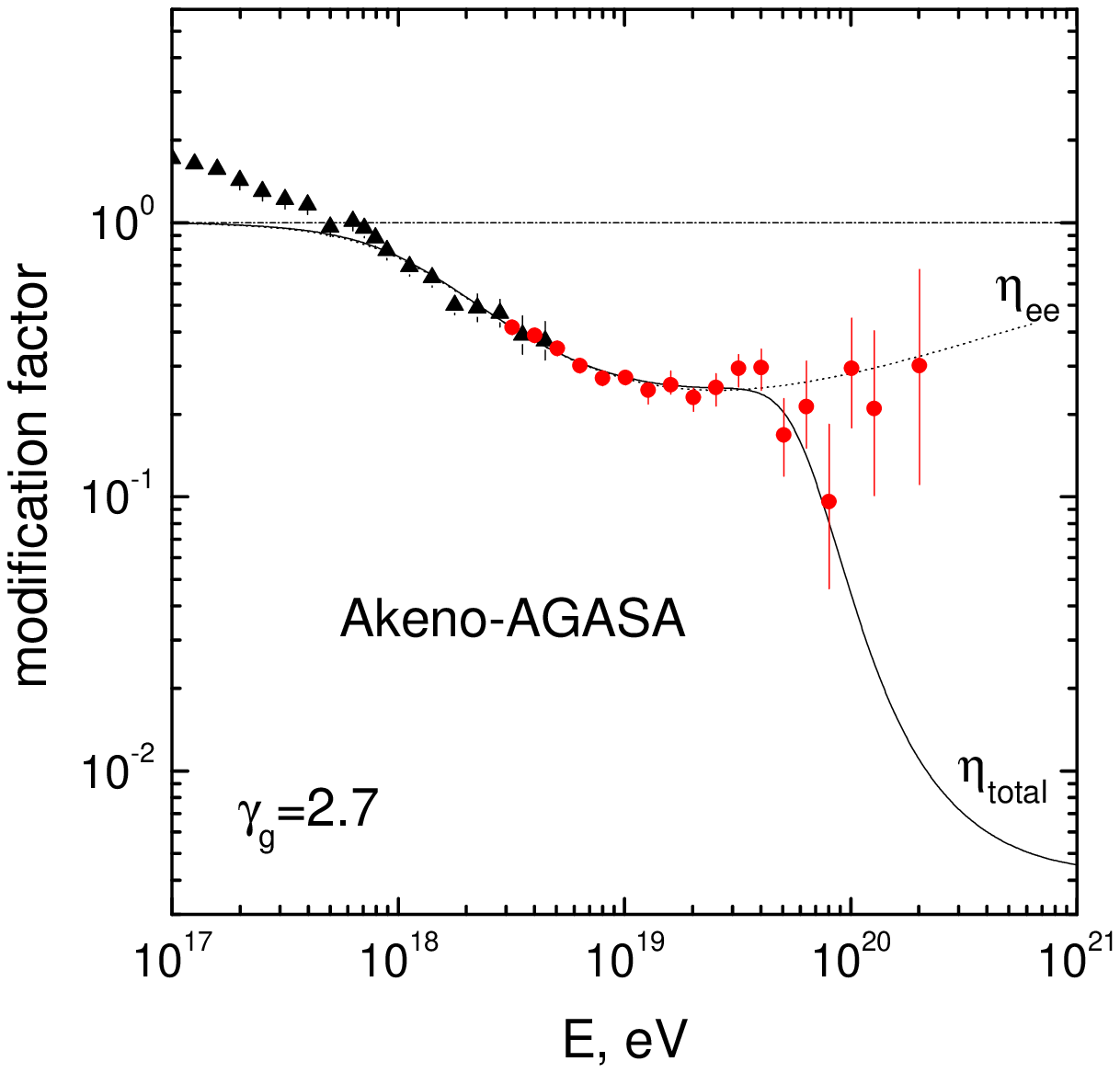}}
\mbox{\includegraphics[height=5cm,width=7.3cm]{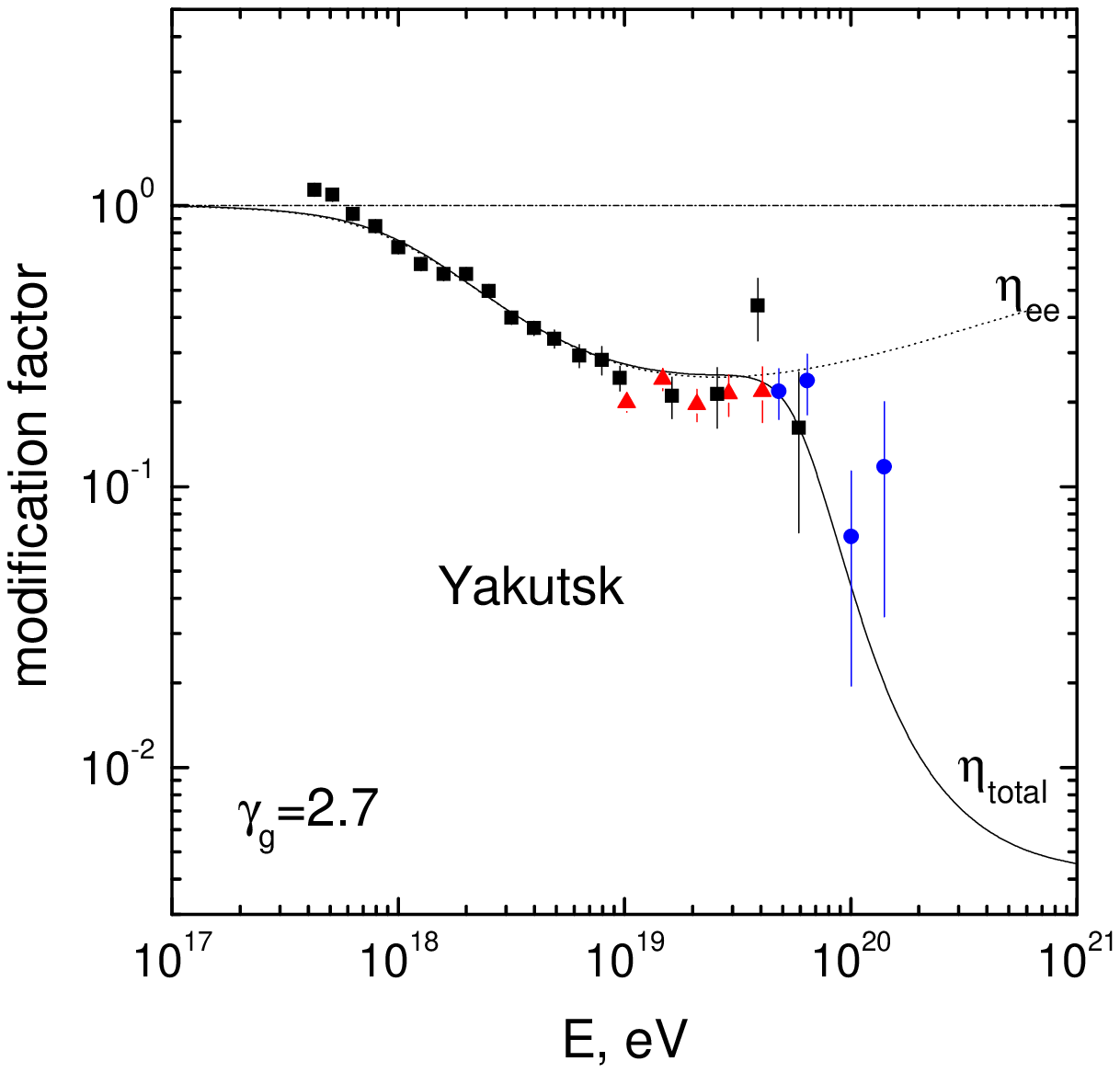}}

\bigskip 
\mbox{\includegraphics[height=5cm,width=7.5cm]{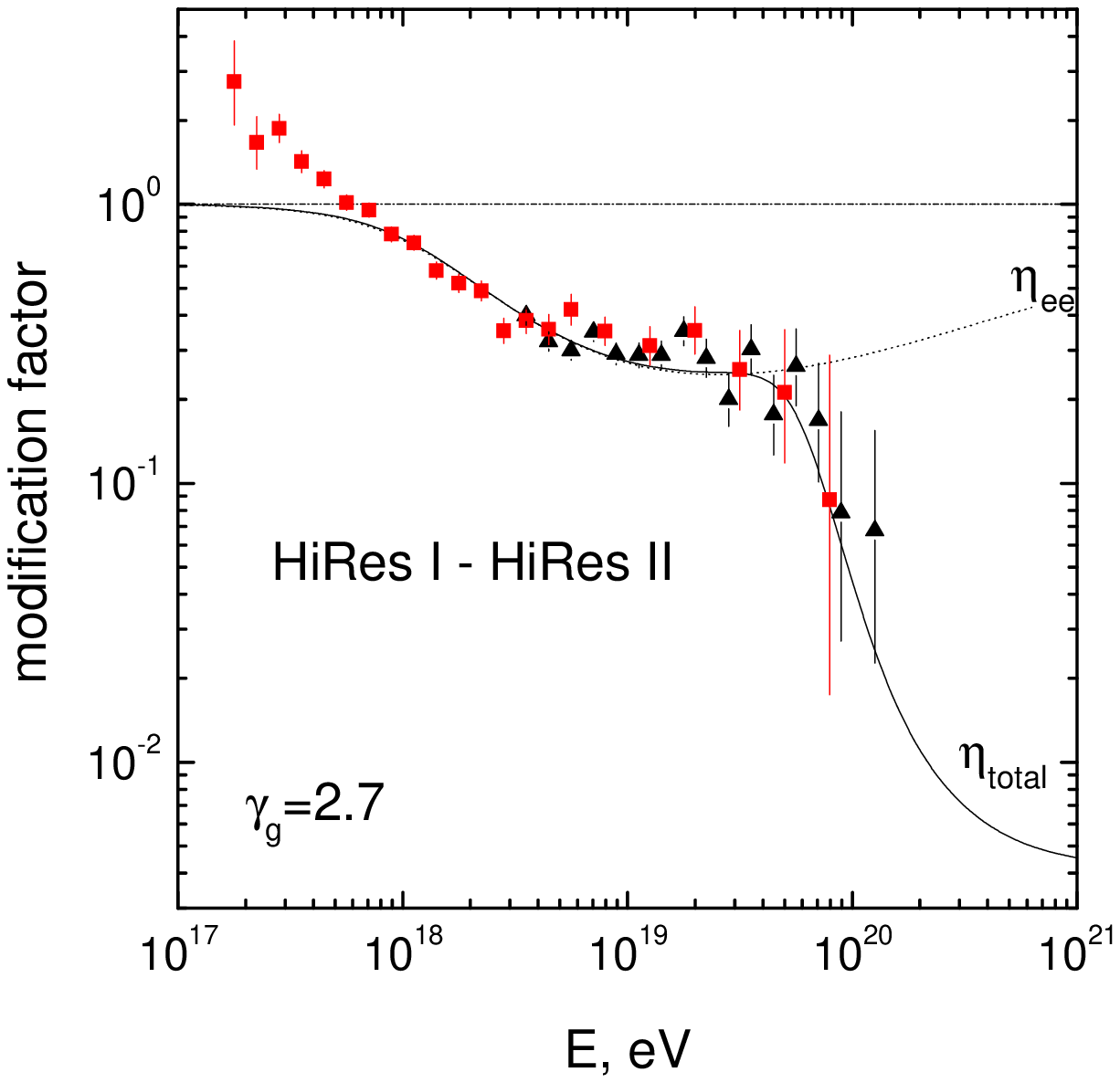}}
\mbox{\includegraphics[height=5cm,width=7.3cm]{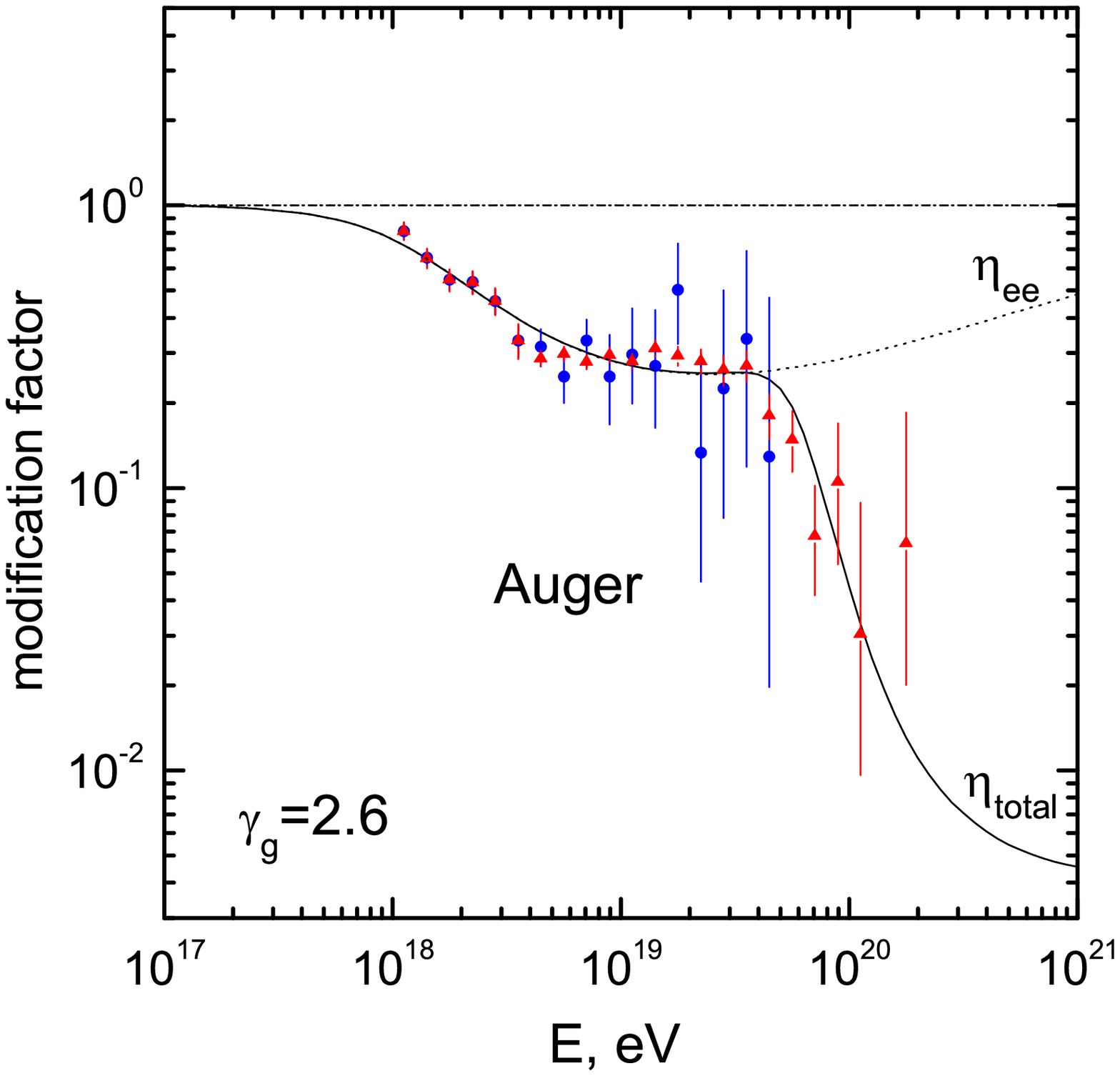}}
\end{center}
\caption{Theoretical pair-production dip and GZK cutoff  in comparison with
the observational data for non-evolutionary models with generation 
index $\gamma_g = 2.6 - 2.7$. The data of HiRes and Auger detectors 
show steepening of the spectrum consistent with the GZK cutoff. The excess
of experimental modification factor over $\eta=1$ at 
$E < 1\times 10^{18}$~ eV in the AGASA and HiRes data evidences for a 
new component, which is given by galactic cosmic rays.
}
\label{fig:dips}
\end{figure*}

In the discussion of the dip above we have not included the cosmological 
evolution, because the evolution is described as minimum by 
two additional free parameters, and agreement of the dip with
observations could look less convincing. The effect of evolution 
was included in calculations \cite{BGG-prd} under assumption 
that AGN are the sources of UHECR \cite{BGG-AGN}. Using the evolution
of UHECR sources close to that observed for AGN, the dip was found 
in good agreement with observations. 
\section{GZK cutoff}
\label{sec:GZK}
From Fig.~\ref{fig:dips} one can see that beginning of the GZK 
cutoff at energy up to $E \sim 8\times 10^{19}$~eV is consistent with
all data including that of AGASA (the events in three  highest energy bins
are the problem of UHE experimental cosmic ray physics and maybe these
events are initiated with small probability by lower energy particles).  
The data of Auger \cite{Auger} and especially HiRes \cite{Hires} agree with 
presence of GZK cutoff. However, low statistics and a possibility of 
imitation of the observed steepening by some other effect, e.g. by 
``acceleration cutoff'', precludes one from making the final
conclusion. For the Auger data we use in Fig.~\ref{fig:dips} the data 
presented at 30th ICRC \cite{Auger}, the data of the last publication  
\cite{Auger-prl} have worse agreement with the predicted cutoff spectrum.  

Recently HiRes collaboration obtained \cite{Hires} numerical
confirmation that steepening seen in Fig.~\ref{fig:dips} is the
GZK cutoff indeed. In the integral spectrum the GZK cutoff
is characterized by energy $E_{1/2}$, where calculated spectrum
$J(>E)$ becomes half of power-law extrapolation spectrum $K
E^{-\gamma}$ from low energies. As calculations \cite{BG88} show this
energy is $E_{1/2} = 10^{19.72}$~eV for a wide range of generation
indices from 2.1 to 2.8. HiRes collaboration found $E_{1/2} =
10^{19.73 \pm 0.07}$~eV in a good agreement with the theoretical
prediction. In Fig.~\ref{fig:E-half} we reproduce the HiRes graph
from which $E_{1/2}$ was determined. The plotted value is given by
ratio of measured flux $J(>E)$ and its power-law approximation
$KE^{-\gamma}$. Extrapolation of this ratio to the higher energies
is given by unity, while intersection of measured ratio with horizontal
line 1/2 gives $E_{1/2}$.
\begin{figure*}[t]
\begin{center}
\mbox{\includegraphics[width=0.65\textwidth,height=5.8cm]{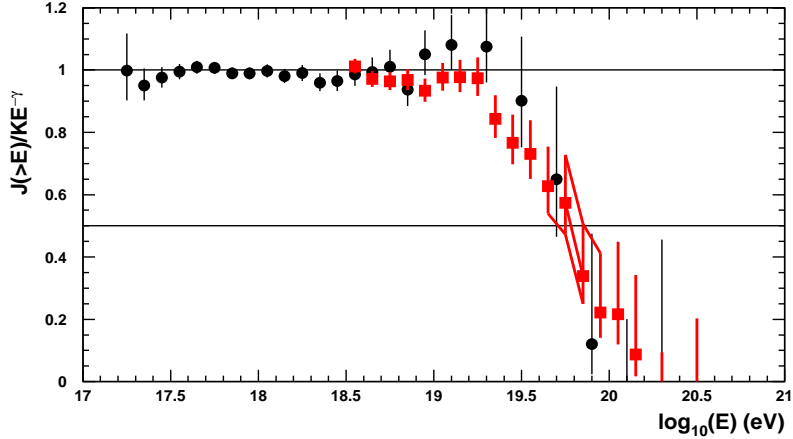}}
\end{center}
\caption{$E_{1/2}$ as numerical characteristic of the GZK cutoff in
  the integral HiRes spectrum (see text).
}
\label{fig:E-half}
\end{figure*}
\noindent
\section{Calibration of detectors with help of dip and GZK
cutoff}
The fluxes measured by Yakutsk, AGASA, HiRes and Auger detectors 
do not agree in the absolute fluxes (see Fig.~\ref{fig:data}). 
This 
discrepancy to the large extent is caused by comparison of the values
$E^3J(E)$ and thus accuracy of energy determination affects strongly the
observed contradiction. In the works \cite{BGG-prd,Aloisioetal} 
the energy calibration performed  with help of the dip 
results in good agreement between the absolute fluxes of all detectors. 
Here we use another approach for calibration based 
on both features, dip and GZK cutoff \cite{taup07}. Since energies as 
measured  by  HiRes fit well the both features and especially the GZK 
numerical characteristic $E_{1/2}$, we assume that HiRes energy scale
is correct and the energies of all other detectors must be shifted by 
factor $\lambda$ to reach the best agreement in fluxes. This procedure 
gives values of $\lambda$ equal to 1.2, 0.75, 0.83, and 0.625 for Auger, 
AGASA, Akeno and Yakutsk, respectively. This calibration provides good 
$\chi^2$ for the dip shape, though does not correspond to the minimum 
$\chi^2$. However, it describes better the dip and beginning of GZK
cutoff taken together. The fluxes after this
energy calibration are shown in Fig.~\ref{fig:calibration} (right
panel). Note, in particular the good agreement of Auger and 
HiRes fluxes, for which the Auger energy scale is increased by 20\%
allowed by Auger systematic energy error. 
\begin{figure*}[t]
\begin{center}
\mbox{\includegraphics[width=0.75\textwidth]{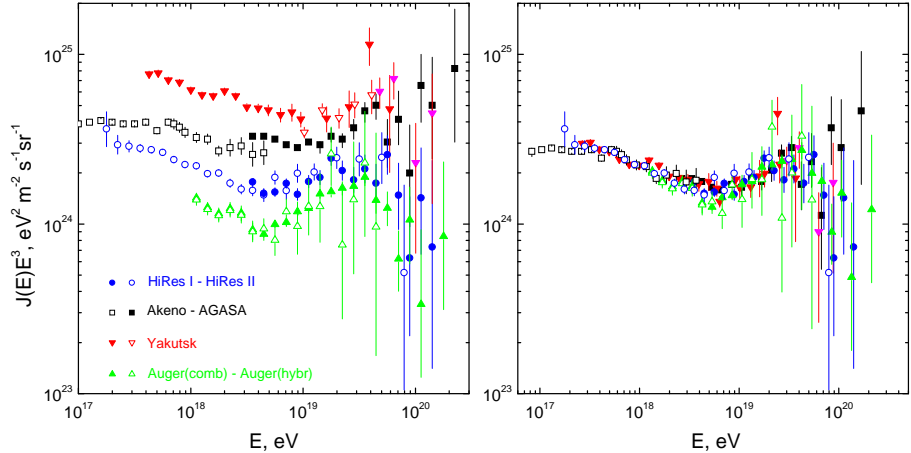}}
\end{center}
\caption{The spectra and fluxes measured by Yakutsk, AGASA,
HiRes and Auger before (left panel) and after (right panel)
energy calibration. The figure is taken from \cite{taup07}.
}
\label{fig:calibration}
\end{figure*}
\section{Mass composition}
In this paper we consider the signatures of UHE protons. 
Their observational confirmations indicate the dominance of the 
proton component in the primary flux. This conclusion must be 
also supported by the direct measurement of the mass composition.

The mass composition measured in two biggest experiments, HiRes and Auger, 
is contradicting. While Hires data \cite{Hires-comp} favour at 
$E \gsim 1\times 10^{18}$~eV the proton-dominated composition, Auger  
observatory claims the mixed-nuclei composition at the same energies
\cite{Auger-mass}.

How the spectral data, in particular dip and GZK cutoff, can be
explained if the primary flux is dominated by nuclei? 

The dip can be produced by transition from the steep galactic component 
at low energies to the flat extragalactic component at higher energies.  
This two-component model was first proposed by Hill and Schramm \cite{HS85} 
in 1985. It was developed in detail in the mixed composition model by 
Allard et al \cite{Allardetal}.
The two spectra, galactic and extragalactic, are equal at the point of
transition $E_{\rm tr} \approx 3\times 10^{18}$~eV. The both
components have mixed nuclei composition. They describe most precisely 
the observed Auger composition, except the highest energy point.  
Energy spectrum in the region of the dip is mostly taken ad hoc. 
Assuming that the two-component dip  fits 
precisely the observed spectrum, one must answer the question why 
spectrum in this ad hoc model is precisely the 
same as in the pair-production dip.    
\section{The sources}
The sources of UHECR must satisfy at least two basic conditions: 
acceleration to energy $E \sim 10^{20} - 10^{21}$~eV and emissivity 
in cosmic rays $\mathcal{L} \gsim 10^{46}$~erg Mpc$^{-3}$ yr$^{-1}$, 
where in terms of space density of the sources $n_s$ and their
luminosity $L_{\rm cr}$ the emissivity is $\mathcal{L}=n_s L_{cr}$. 
These two conditions are satisfied by AGN and GRBs.

The sources of GRBs are assumed to be hypernovae, where particles are
accelerated by the shocks in jet or in external shock 
(see \cite{Dermer1} for description and references). The protons can
reach energies up to $E_{\rm max} \sim 1\times 10^{21}$~eV, but there
is a problem with the total energy output \cite{BGG-prd}. The interesting
possibility considered in \cite{Dermer2} is that hypernovae produce 
both galactic and extragalactic cosmic rays. Galactic cosmic rays are 
produced by a single hypernova in Milky Way exploded $2\times 10^5$~yr 
ago. The particles with energies $E > 10^{18}$~eV escaped from Milky Way
and the observed flux at these energies are produced by extragalactic
hypernova. 

AGN are one of the best candidates for UHECR sources, as far as 
acceleration and energetics is concerned. The dip, calculated 
with cosmological evolution of AGN, as observed in X-rays,
agrees well with measured spectra \cite{BGG-prd}. 

The plausible candidates for sources of UHECR are Fanaroff-Riley type 1
(FR1) radiogalaxies. They are AGN with short jets where acceleration
most probably occurs.  FR1 with jets directed to observer compose 
the population of BL Lacs. There are indications to correlations of 
BL Lacs with 
UHECR in AGASA \cite{bllac-AGASA} and HiRes \cite{bllac-Hires} data,  
but these correlations are absent in Auger observations. 
\section{Conclusions}
Extragalactic UHE protons propagating through CMB acquire two features
in the energy spectrum: pair-production dip and GZK cutoff. The 
pair-production dip is a faint spectral feature located at energies 
$1 - 40$~EeV. The large statistics of observations provide the accurate  
measurement of the dip energy spectrum. Its shape agrees very well with 
the theoretical prediction. The part of the GZK feature, up to 
80~EeV, is seen in all experimental data including that of AGASA.  
The HiRes spectrum agrees with the predicted GZK spectrum up to 
100~EeV within the limited statistics of observations. The measured 
characteristic of the GZK cutoff in the integral spectrum of HiRes, 
$E_{1/2} = 53$~EeV, coincides with the theoretical prediction. 
Spectrum measured by Auger detector, as it was presented at 30th 
ICRC, does not contradict the calculated spectrum, too.  

The good agreement of these two spectral features
with theoretical prediction implies that primary particles are mostly 
extragalactic protons. The direct measurements of the mass composition 
by $X_{max}$ method is contradicting. HiRes data favour the 
proton-dominated composition, Auger data -- the mixed-nuclei composition.

The absolute fluxes in terms of $E^3J(E)$ measured by HiRes, Auger,
AGASA and Yakutsk detectors differ much from each other. Difference
in energy scale in the experiments is responsible for this
discrepancy. Assuming that energy scale of the HiRes detector is correct   
and shifting energies in other experiments by factors $\lambda$,
different for each experiment, the agreement of all data is obtained.   
\section*{Acknowledgements} 
I thank my collaborators R.~Aloisio, A.~Gazizov and S.~Grigorieva 
for discussions. This work was supported in part by contract ASI-INAF
1/088/06/0 for theoretical studies in High Energy Astrophysics.

\end{document}